\documentclass[preprint,superscriptaddress,prl,showpacs]{revtex4}
\usepackage{amssymb,amsmath,graphicx,amscd,xcolor}

\begin{document}

\title{Vacuum Lightcone Fluctuations in a Dielectric}

\author{C. H. G. Bessa}
\email{carlos@cosmos.phy.tufts.edu}
\affiliation{Institute of Cosmology, Department of Physics and Astronomy \\
Tufts University, Medford, Massachusetts 02155, USA}

\author{V. A. De Lorenci}
\email{delorenci@unifei.edu.br}
\affiliation{Instituto de F\'{\i}sica e Qu\'{\i}mica,    Universidade Federal de Itajub\'a \\
 Itajub\'a, Minas Gerais 37500-903, Brazil}
\affiliation{Institute of Cosmology, Department of Physics and Astronomy \\
Tufts University, Medford, Massachusetts 02155, USA}

\author{L. H. Ford}
\email{ford@cosmos.phy.tufts.edu}
\affiliation{Institute of Cosmology, Department of Physics and Astronomy \\
Tufts University, Medford, Massachusetts 02155, USA}

\author{N. F. Svaiter}
\email{nfuxsvai@cbpf.br}
\affiliation{Centro Brasileiro de Pesquisas F\'{\i}sicas \\
Rua Dr. Xavier Sigaud 150, 22290-180  Rio de Janeiro, RJ, Brazil}

\begin{abstract}
A model for observable effects of electromagnetic vacuum fluctuations is presented. The model involves
a probe pulse which traverses a slab of nonlinear optical material with a nonzero second order 
polarizability. We argue that the pulse interacts with the ambient vacuum fluctuations of other
modes of the quantized electric field, and these vacuum fluctuations cause variations in the flight 
time of the pulse through the material. The geometry of the slab of material defines a sampling function
for the quantized electric field, which in turn determines that vacuum modes whose wavelengths are of the
order of the thickness of the slab give the dominant contribution. Some numerical estimates are made, which
indicate that fractional fluctuations in flight time of the order of $10^{-9}$ are possible in realistic situations.
The model presented here is both an illustration of a physical effect of vacuum fluctuations, and an
analog model for the lightcone fluctuations predicted by quantum gravity.   
 \end{abstract}
		
\maketitle
\baselineskip=13pt	

Quantum fluctuations of the electromagnetic field are responsible for several observed
phenomena, including the Lamb shift and the Casimir effect. However, there is still
debate about the reality of vacuum fluctuations~\cite{Boddy}.  Here we explore the viewpoint that vacuum fluctuations
can be just as real as thermal fluctuations, but  are often not noticed because of strong anticorrelations.
The anticorrelations prevent an electric charge from undergoing observable Brownian motion
in the vacuum state. The charge can temporarily acquire energy from a vacuum electric field
fluctuation, but this energy will be taken away by an  anticorrelated fluctuation on a time scale
consistent with the energy-time uncertainty principle. This viewpoint is supported by calculations
in models where the cancellation is upset by a time dependent background~\cite{YF04,BF09,PF11,PF14}.
We will here construct a model without an explicit external time dependent background, 
but where vacuum electric field fluctuations have a clear, and potentially observable, physical effect.

Let $E$ be a Cartesian component of the quantized electric field operator. Vacuum expectation values of
even powers of $E$  are divergent due to the contribution of high frequency modes. This
divergence may be removed by replacing $E$ by its time average with a suitable sampling function,
or test function, $f_\tau(t)$, where $\tau$ is the characteristic width of the function. Let
\begin{equation}
\bar{E} = \int_{-\infty}^\infty  E(t) \, f_\tau(t) \, dt \,.
\label{eq:t-ave}
\end{equation}
Here $E(t)$ is the field operator at any fixed space point, and 
\begin{equation}
\int_{-\infty}^\infty f_\tau(t) \, dt =1 \,.
\label{eq:fnorm}
\end{equation} 
The moments of $\bar{E}$ are finite and those of a Gaussian distribution, determined by the second
moment
\begin{equation}
\langle 0| \bar{E}^2 |0 \rangle = \frac{a}{\tau^4} \,,
\label{eq:E2}
\end{equation} 
where the numerical constant $a$ depends upon the choice of sampling function. 
(Lorentz-Heaviside units with $c=\hbar =1$ will be used here, except as otherwise noted.) For the
case of a Lorentzian function,
 \begin{equation}
f_\tau(t) = \frac{\tau}{\pi(t^2 + \tau^2)} \,,
\label{eq:Lor}
\end{equation}
we have $a=1/\pi^2$. Modes whose
period is of order $\tau$ give the dominant contribution here, with the contribution of shorter
wavelength suppressed by the time averaging. In rigorous treatments of quantum field theory,
test functions, usually with compact support, are used to define well-behaved operators. 
See, for example, Ref.~\cite{PCT}.   However,
this use of test functions is purely formal, and no physical interpretation is made. One of the purposes 
of this letter will be to provide an example where the function  $ f_\tau(t)$ has a clear meaning defined 
by the physical system of interest.

Our model will involve light propagation in a nonlinear material. Related models were presented
in Ref.~\cite{BDF14}, as an analog model for semiclassical gravity, and in Ref.~\cite{flms13}, as a model
for the lightcone fluctuations expected in quantum gravity~\cite{Pauli,Deser,DeWitt,F95,FS96,FS97}.  
In a nonlinear material,   the polarization is a nonlinear function~\cite{Boyd}  of the electric field:
\begin{equation}
P_i = \left(\chi_{ij}^{(1)} E_{j} + \chi_{ijk}^{(2)} E_{j}E_{k} + \cdots \right) \,,
\label{eq:pol}
\end{equation}
where repeated indices are summed upon. Here $\chi_{ij}^{(1)}$ and $\chi_{ijk}^{(2)}$ are the first and
second order susceptibility tensors, respectively. The second and higher order susceptibilities lead to 
a nonlinear wave equation for the electric field.  We assume that the total electric field may be written 
as the sum of a background field ${\bf E}_0$ and a smaller but more rapidly varying probe field  ${\bf E}_1$,
\begin{equation}
{\mathbf E} = {\mathbf E}_0 + {\mathbf E}_1 \,.
 \end{equation} 
 Both ${\bf E}_0$ and ${\bf E}_1$ satisfy nonlinear equations, with a coupling term between them.
 Here the background field describes the vacuum modes of the quantized electric field, and will be
 approximated as a linear field in an approximately isotropic materials, so we set 
 \begin{equation}
 \chi_{ij}^{(1)} \approx \delta_{ij}\,  \chi^{(1)}\,.
 \label{eq:iso}
\end{equation}
We take the probe field to be polarized in the $z$-direction, ${\mathbf E}_1 = E_1(t,x,y){\bf \hat{z}}$,
 and ignore its self-coupling. Its linearized wave equation may be written as~\cite{flms13}
 \begin{equation}
\frac{\partial^2 E_1}{\partial x^2} + \frac{\partial^2 E_1}{\partial y^2} - \frac{1}{v^2}\left(1  +
 2 \epsilon_{1}\right) 
\frac{\partial^2 E_{1}}{\partial t^2}  = 0\,.
\label{eq:we1}
\end{equation}
Here  
\begin{equation}
 v = \frac{1}{\sqrt{1 + \chi^{(1)}}} = \frac{1}{n_p} \, ,
 \label{speed}
 \end{equation}
is the speed of a probe pulse in the medium with index of refraction $n_p$ due to linear effects, and
\begin{equation} 
\epsilon_1 =\gamma_j \,E_0^j \,,
\end{equation}
with
\begin{equation}
\gamma_j  = \frac{1}{n_p^2}\, \left(\frac{\chi_{zzj}^{(2)} + \chi_{zjz}^{(2)}}{2}\right) \,.
\label{eq:gamma}
 \end{equation}
Thus the phase velocity of the probe field will be given by, assuming $|\epsilon_1| \ll1$,
\begin{equation}
v_{ph} = \frac{v}{\sqrt{1 + 2\epsilon_1 }} \approx v (1 - \epsilon_1) \,, 
\end{equation}
which depends upon the value of the background field, ${\bf E}_0$. If we form wavepackets
over a frequency range where dispersion is small, then this is also the group velocity of the
packets. For a slab of material with thickness $d$, the flight time of a pulse propagating in the
$x$-direction will be $d/v$ without the background field, and approximately
\begin{equation}
t_d =  n_p \int_0^{d}   (1 + \epsilon_1)\, dx
\label{eq:td}
\end{equation}
with the background field.

Fluctuations of the background field cause fluctuations in the speed of the probe field. Here we take
these to be the vacuum fluctuations of the quantized electric field. We will work in an approximation in 
which the various contributing modes of the quantized field propagate at approximately the same
speed, and hence experience a frequency independent and isotropic index of refraction $n_b$,
different from $n_p$. We will
need the electric field correlation functions in such a material, which may be  obtained from
the corresponding correlation functions in empty space by a simple argument.  First consider
the standing modes in a cavity of fixed geometry, and assume periodic boundary conditions. The spatial
part of a mode function is independent of $n_b$, and will be proportional to ${\rm e}^{i {\bf k \cdot x}}$.
The temporal part will oscillate at an angular frequency $\omega = k/n_b$. The Faraday law, 
${\bf \nabla \times E} = -{\bf \dot{B}}$, tells us that the magnitudes of the electric and magnetic fields
of a mode are related by $B = (k/\omega) E = n_b \, E$. We require that the zero point energy of
a given mode be  $\omega/2$ as $n_b$ varies, which implies
\begin{equation}
\frac{1}{2}\, \omega = \frac{k}{2 n_b} = \frac{1}{2} \int d^3x ( n_b^2 E^2 + B^2)\,.
\end{equation}
As a result, we have
\begin{equation}
E \propto \frac{1}{n_b^{3/2}}\,, \qquad B \propto \frac{1}{n_b^{1/2}}\,,
\end{equation}
which agrees with Eqs.~(1.31a) and (1.32) of Ref.~\cite{GL92}.
This shows that the net effect on an electric field
correlation function is an overall factor of $1/n_b^3$ and a modification of the time dependence by
$ t \rightarrow   t/n_b$, with no effect on the space dependence. This result may also be obtained 
from Eqs.~(5) and (6) in Ref.~\cite{BHL92}. Take the spatial
separation to be in the $x$-direction, in which case the correlation functions become
 \begin{equation}
\langle E^x(x)E^x(x')\rangle = \frac{1}{\pi^2\, n_b^3\, \left[(\Delta x)^2 - (\Delta t)^2/n_b^2\right]^2}\,,
\label{eq:Exx}
\end{equation}
and
\begin{equation}
\langle E^y(x)E^y(x')\rangle= \langle E^z(x)E^z(x')\rangle = 
\frac{(\Delta x)^2 + (\Delta t)^2/n_b^2}{\pi^2 \, n_b^3\, \left[ (\Delta t)^2/n_b^2 - (\Delta x)^2\right]^3}\,.
\label{eq:Eyy}
\end{equation}
Here $\Delta x = x-x'$ and $\Delta t = t -t' -i \epsilon$, and $\epsilon > 0$ makes the mode sums absolutely
convergent and defines the location of the lightcone singularity. 
Note that the effective lightcone is given by the line $ t = n_b\,  x$, with $n_b >1$.

\begin{figure}
 \centering
 \includegraphics[scale=0.8]{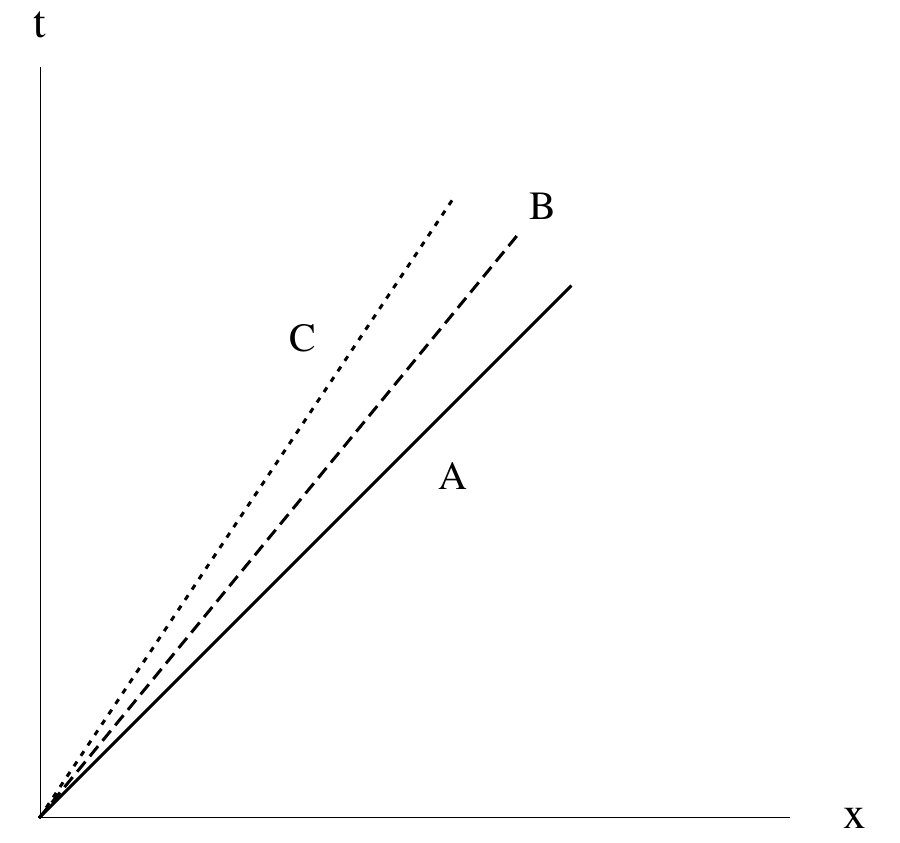}
 \caption{Here $A$ is the true lightcone ($t =x$), $B$ is the effective lightcone  ($t = n_b x$),
 and $C$ is the worldline of the probe pulse ($t = n_p x$).}
 \label{fig:lightcones}
 \end{figure}

We wish to consider the case where the probe pulse is in a higher frequency band than the dominant
vacuum modes, and has an index of refraction of $n_p > n_b$. Thus the probe pulse travels on
a worldline which lies inside the effective lightcone, which is turn inside the true lightcone, as illustrated 
in Fig.~\ref{fig:lightcones}. The travel time of a pulse through a slab of material is given by an integral
of the form of that in Eq.~\eqref{eq:td}. However, this travel time undergoes fluctuations around a mean value
of $\langle t_d \rangle = d/v$, and with a variance of
\begin{equation}
(\delta t)^2 = \langle t_d^2 \rangle -  \langle t_d \rangle^2 = n_p^2 \int_0^d dx \int_0^d dx' \,
 \langle \epsilon_1(x) \epsilon_1(x') \rangle \,.
 \label{eq:Dt}
\end{equation}
The correlation function $ \langle \epsilon_1(t) \epsilon_1(t') \rangle$ may be expressed in terms of
the electric field correlation functions given in Eqs.~\eqref{eq:Exx} and \eqref{eq:Eyy}. The integration
in Eq.~\eqref{eq:Dt} is along the path of the probe pulse, defined by $x = t/n_p$, and illustrated in
Fig.~\ref{fig:lightcones}. 

\begin{figure}
 \centering
 \includegraphics[scale=0.8]{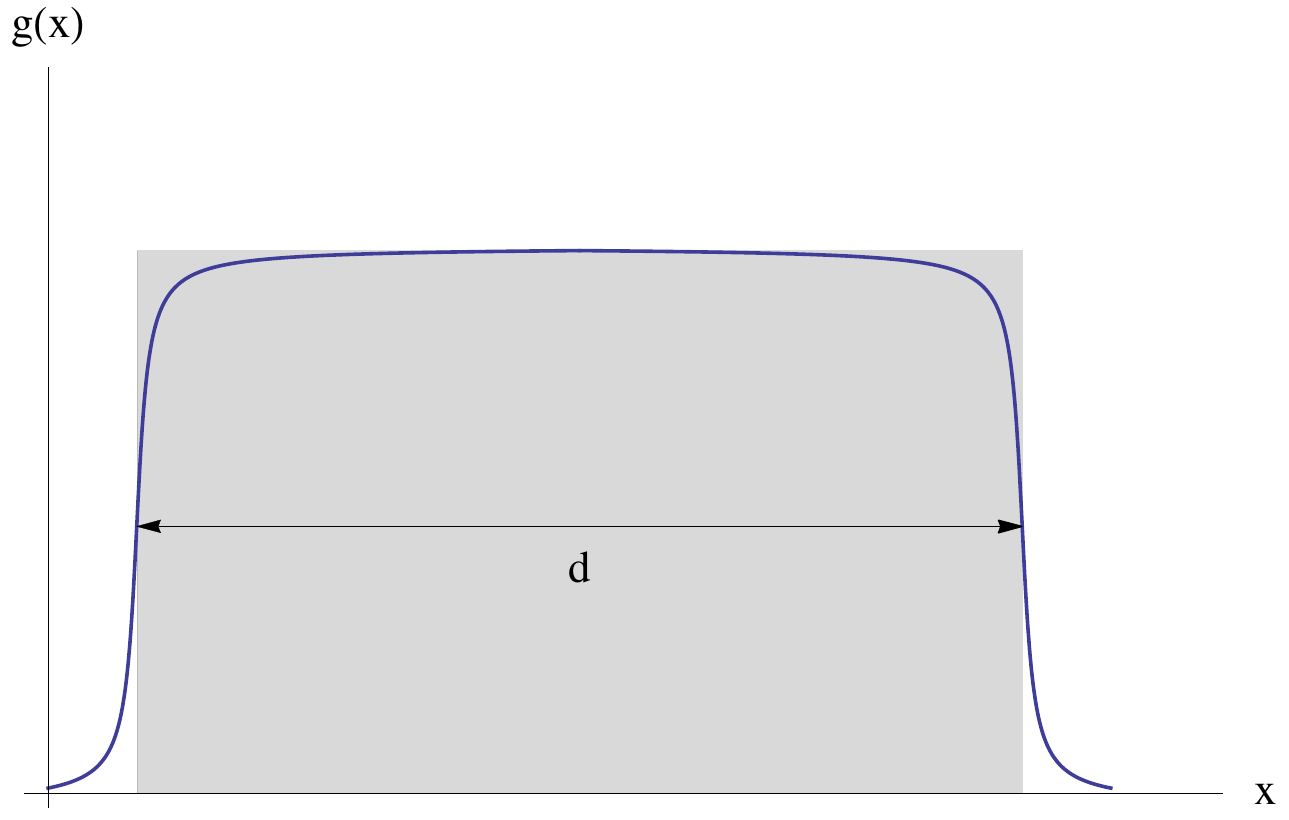}
 \caption{The profile function $g(x)$ is illustrated.}
 \label{fig:profile}
 \end{figure}

The integrals of the electric field correlation functions will be well defined provided there is a
sampling function which falls smoothly to zero at both ends of the integration range. In the present
context, such a function can be provided by the geometry of the slab of nonlinear material. Suppose 
that the density of this material is tapered gradually at both ends, so that $\chi_{ijk}^{(2)}$ can be
replaced by  $\chi_{ijk}^{(2)}\, g(x)$, where $g(x)$ is a profile function of width $d$, and 
$\int_{-\infty}^\infty dx\, g(x) =d$,  illustrated in  Fig.~{\ref{fig:profile}}.
This profile function, along with the worldline of the probe pulse, define a temporal sampling function,
$f_\tau(t) = g(t/n_p)/\tau$, whose characteristic width is $\tau = n_p \,d$. We take the normalization of  
$f_\tau(t)$ to be defined by Eq.~\eqref{eq:fnorm}. The effect of the profile
function is to insert a factor of $g(x)\, g(x')$ in the integrand of Eq.~\eqref{eq:Dt}, and to extend the range
of integration to all $x$

The fractional variance in flight time may be expressed as
\begin{equation}
\delta^2 = \frac{(\delta t)^2}{ \langle t_d \rangle^2} = \int_{-\infty}^\infty dt \,dt' \, f_\tau(t)  f_\tau(t')\;
\gamma_j \gamma_k \,\langle E^j(t)E^k(t')\rangle \,.
\end{equation}
If we use Eqs.~\eqref{eq:Exx} and \eqref{eq:Eyy}, and the fact that here $\langle E^j(t)E^k(t')\rangle =0$
for $j \not= k$, we find
\begin{equation}
\delta^2 = \frac{C\,n_b}{\pi^2 (n_p^2 -n_b^2)^3\, d^4} \, [(n_p^2-n_b^2) \gamma_1^2
+ (n_p^2+ n_b^2) (\gamma_2^2 + \gamma_3^2)] \,.
\label{eq:del2}
\end{equation}
Here we have used $d = \tau/n_p$ and the result
\begin{equation}
\int_{-\infty}^\infty dt \,dt' \, \frac{f_\tau(t)  f_\tau(t')}{(t-t' -i \epsilon)^4} = \frac{C}{\tau^4}\,. 
\end{equation}
Here $C$ is a dimensionless constant whose value depends upon the choice of the sampling
function $f_\tau(t)$. If this function is a Lorentzian, Eq.~\eqref{eq:Lor}, then $C = 1/16$.

The root mean square of the fractional flight time variation is
\begin{equation}
\delta_{\rm rms} = \sqrt{\delta^2} \propto \frac{ \chi^{(2)}}{d^2}\,,
\end{equation}
where $\chi^{(2)}$ is a component of the $\chi_{ijk}^{(2)}$ tensor, which has typical values
of the order of $10^{-12} {\rm m/V}$ in SI units. The dimensionless ratio, ${\chi^{(2)}}/{d^2}$,
may be expressed as
\begin{equation}
\frac{\chi^{(2)}}{d^2} = 1.9\times10^{-8}\left[\frac{\chi^{(2)}}{10^{-12}m/V}\right]\left(\frac{1\mu m}{d}\right)^2\,.
\end{equation}

A specific example of a material is that of Cadmium selenide (CdSe), which has 
$\chi_{zzz}^{(2)} \approx 1.1 \times 10^{-10}m/V$ at a wavelength of $10.6 \mu m$ \cite{Penzkofer},
and indices of refraction of $n_p =2.54$ at $\lambda = \lambda_p =1.06 \mu m$ and of
$n_b =2.43$ (ordinary ray)  and $n_b =2.44$ (extraordinary ray)  at 
$\lambda = \lambda_b =10.6 \mu m$ ~\cite{HB71,Bh76}. The nearly equal values of $n_b$ for the
ordinary and extraordinary rays justifies the isotropy assumption in Eq.~\eqref{eq:iso}.
The other components of $\chi_{ijk}^{(2)}$
which appear in Eq.~(\ref{eq:del2}) vanish. The crystal lattice of this material has hexagonal
symmetry, and hence a single axis of rotational symmetry, which is taken to be the $z$-axis.
 With $C=1/16$, we find the estimate
\begin{equation}
\delta_{\rm rms} = 3.6 \times 10^{-9} \, \left(\frac{10 \mu m}{d} \right)^2 \,.
\label{eq:del2b}
\end{equation}
This is an estimate of the fractional variation in flight times of wavepackets peaked at a mean wavelength
of $\lambda_p \approx 1 \mu m$ traversing a distance $d$. Clearly the spatial spread in the wavepackets
is relevant here. The bandwidth in angular frequency cannot be more that about 
$\Delta \omega \approx 2 \pi/\lambda_p$ with a corresponding spatial spread of 
$\Delta x = 1/\Delta \omega = \lambda_p/2 \pi$. With $d = 10 \mu m$, a single pulse could measure a
flight time to an accuracy of about $2 \times 10^{-2}$ at best. However, averaging over a very large number
of pulses might lead to an accurate determination of $\delta_{\rm rms}$, which is an uncertainty in flight
time due to vacuum fluctuations. If it is possible to do experiments with smaller slabs, and correspondingly
shorter wavelengths, then the effect could be larger, due to the $1/d^2$ dependence of $\delta_{\rm rms}$.   

Here we should comment on the assumption that $|E_1| \ll |E_0|$ which was used in deriving 
Eq.~\eqref{eq:we1}.  Despite the fact that the probe field modes are of higher frequency than the relevant
vacuum modes, this condition can still be satisfied for highly collimated probe beams with narrow bandwidths.
For the sake of an estimate, we ignore the indices of refraction, and write 
$E_1^2 \approx |z|^2\, \lambda_p^{-4}\, (\Delta \omega/\omega)\, \Delta \Omega$. Here $|z|^2$ is the mean 
number of photons per mode, $\Delta \omega/\omega$ is the fractional bandwidth, and $\Delta \Omega$
is the solid angle subtended by the probe beam. This is to be compared with Eq.~\eqref{eq:E2},
$E_0^2 = a/\tau^4$. Even though $\lambda_p < \tau$, we can still have $|E_1| \ll |E_0|$ if
$\Delta \omega/\omega \ll 1$ or $\Delta \Omega \ll 1$. This arises because fewer modes contribute
to $|E_1|$ than to  $|E_0|$.

The model may be extended to include the effects of squared electric field fluctuations. Here we give
a brief account, and a more detailed version will be presented elsewhere~\cite{next}. If we include
third-order polarizability terms in Eq.~\eqref{eq:pol}, then the wave equation for the probe field,
Eq.~\eqref{eq:we1}, will contain a term $\epsilon_2 \propto \chi^{(3)} (E_0)^2$, ~\cite{BDF14,flms13},
where  $\chi^{(3)}$ is a component of the third-order polarizability tensor, $\chi_{ijkl}^{(3)}$, and
$E_0$ is a component of the background electric field. Now the probe pulse is sampling the squared
electric field operator on a timescale $\tau$. Let $E^2$ denote the normal order of the squared electric
field operator, and $\bar{E^2}$ denote its time average, defined as in Eq.~\eqref{eq:t-ave}. Now the
analog of Eq.~\eqref{eq:E2} is
\begin{equation}
\langle 0| (\bar{E^2})^2 |0 \rangle = \frac{b}{\tau^8} \,,
\label{eq:E4}
\end{equation} 
where $b$ is a dimensionless constant determined by the sampling function. We can make an
estimate of the magnitude of the squared electric field fluctuations by simply replacing 
$\chi^{(2)} \rightarrow \chi^{(3)}$ and $d \rightarrow d^2$ in our previous results. Typical
values of $\chi^{(3)}$ are of order $10^{-18} m^2/V^2$ in SI units. Thus our estimate for the
fractional variation in flight time becomes
\begin{equation}
\delta_{\rm rms} \approx \frac{ \chi^{(3)}}{d^4} = 
4 \times 10^{-14} \,\left( \frac{\chi^{(3)}}{10^{-18}m^2/V^2}\right)\,   \left(\frac{10 \mu m}{d} \right)^4 \,.
\end{equation}   
Thus the effects of $E^2$ fluctuations are typically small compared to those of electric field fluctuations,
but the former grow more rapidly with decreasing $d$ and hence decreasing sampling time.
It is also of interest to note that the probability of large $E^2$ fluctuations can be much larger 
than for electric field fluctuations. In contrast to the Gaussian probability distribution of $E$,
the corresponding distribution for $E^2$ falls approximately as~\cite{FFR12}
\begin{equation}
P(x) \sim x^{-2}\, {\rm e}^{-x^{1/3}}\,, \quad x \gg 1\,,
\end{equation}
where $x= E^2\, (4\pi \tau^2)^2$ is a dimensionless measure of $E^2$, and Lorentzian averaging
is assumed. This means that large fluctuations are less rare than for a Gaussian probability distribution.
The distribution of flight times due to electric field fluctuations, with the variance described by
Eq.~\eqref{eq:del2b}, is Gaussian, while the corresponding distribution due to squared electric field 
fluctuations will be a non-Gaussian distribution of the type discussed in Ref.~\cite{FFR12}, with a long
positive tail.

The model presented here is an analog model for quantum lightcone fluctuations, as was the model of
Ref.~\cite{flms13}. However, in the latter model, the fluctuations were not vacuum fluctuations, but
were generated by a squeezed state of the electromagnetic field. The model described here is an
illustration of the reality of switched vacuum fluctuations, where the details of the switching are given
by the geometry of the slab of nonlinear material. The probe pulse is only sensitive to vacuum fluctuations
occurring in a finite time interval, so a nonzero effect arises, despite the tendency of vacuum fluctuations
to be anticorrelated.

 \begin{acknowledgments}
This work was supported in part by the National Science Foundation under Grant PHY-1205764,
and by the Brazilian research agencies CNPq (245985/2012-3 and 304486/2012-4), FAPEMIG, 
and CAPES under scholarship BEX 18011/12-8.
\end{acknowledgments}


\begin{thebibliography}{99}

\bibitem{Boddy} K. K. Boddy, S. M. Carroll, and J. Pollack, arXiv:1405.0298.

\bibitem{YF04} H. Yu and L. H. Ford,  Phys. Rev. D {\bf 70}, 065009 (2004),
quant-ph/0406122.
 
\bibitem{BF09} C. H. G. Bessa, V. B. Bezerra, and L. H. Ford, J. Math. Phys. {\bf 50},
062501 (2009),  arXiv:0804.1360.

\bibitem{PF11} V. Parkinson and L. H. Ford, Phys. Rev. A {\bf 84}, 062102 (2011),
arXiv:1106.6334.

\bibitem{PF14} V. Parkinson and L. H. Ford, Phys. Rev. D {\bf 89}, 064018 (2014),
arXiv:1311.6422. 

\bibitem{PCT} R. F. Streater and A. S. Wightman, {\it PCT, Spin and Statistics, and All That}, 
(Benjamin, New York, 1964), p 97. 

\bibitem{BDF14} C. H. G. Bessa, V. A. De Lorenci, and  L. H. Ford, Phys. Rev. D {\bf 90},
024036 (2014); arXiv:1402.6285. 

\bibitem{flms13} L. H. Ford, V. A. De Lorenci, G. Menezes, and N. F. Svaiter, 
Ann. Phys. {\bf 329}, 80 (2013);  arXiv:1202.3099.

\bibitem{Pauli} W. Pauli, Helv. Phys. Acta. Suppl. {\bf 4},  69  (1956).
This reference consists of some
remarks made by Pauli during the discussion of a talk by O. Klein at a 1955
conference in Bern, on the 50th anniversary of relativity theory.

\bibitem{Deser}  S. Deser, Rev. Mod Phys. {\bf 29},  417  (1957).

\bibitem{DeWitt}  B. S. DeWitt, Phys. Rev. Lett. {\bf 13},  114  (1964).

\bibitem{F95}  L. H. Ford, Phys. Rev. D {\bf 51},  1692  (1995), arXiv:gr-qc/9410047.

\bibitem{FS96} L. H. Ford and N. F. Svaiter,  Phys. Rev. D {\bf 54},  2640  (1996),
arXiv:gr-qc/9604052. 

\bibitem{FS97}  L. H. Ford and N. F. Svaiter,  Phys. Rev. D  {\bf 56},  2226 (1997),
arXiv:gr-qc/9704050.

\bibitem{Boyd} R. W. Boyd, {\it Nonlinear optics}, 3rd ed. (Academic Press, New York, 2008).

\bibitem{GL92} R. J. Glauber and M. Lewenstein, Phys. Rev. A {\bf 43}, 467 (1992). 

\bibitem{BHL92} S. M. Barnett, B. Hutner, and R. Loudon,  Phys. Rev. Lett. {\bf 68}, 3698 (1992).

 \bibitem{Penzkofer} A. Penzkofer, M. Sch{\"a}ffner, and X. Bao, Opt. and Quantum Electron.
 {\bf 22}, 351 (1990).

\bibitem{HB71} R. L. Herbst and R. L. Byer, App. Phys. Lett. {\bf 19}, 527 (1971). 

\bibitem{Bh76} G. C. Bhar, App. Optics {\bf 15}, 305 (1976).

\bibitem{next} C. H. G. Bessa, V. A. De Lorenci, and  L. H. Ford, manuscript in preparation. 

 \bibitem{FFR12}  C. J. Fewster, L. H. Ford, and T. A. Roman, Phys. Rev. D {\bf 85},
125038 (2012),  arXiv:1204.3570.

\end{thebibliography}
\end{document}